\documentclass[%
 preprint,
nofootinbib,
 amsmath,amssymb,
 aps,
prl,
 longbibliography,
 lengthcheck,%
]{revtex4-1}

\usepackage{graphicx}
\usepackage{dcolumn}
\usepackage{bm}
\usepackage{hyperref}
\usepackage{amsmath,bm}
\usepackage{lineno}

\def\ba{\begin{equation}\begin{array}{c}}
\def\ea{\end{array}\end{equation}}

\renewcommand{\Pi}{\hat P}

\begin{document}

\preprint{APS/123-QED}

\title{
Effect of dephasing on the current\\
 through a periodically driven quantum point contact 
}

\author{Igor Ermakov$^{1,2}$}
\email{ermakov1054@yandex.ru}
\author{Oleg Lychkovskiy$^{2,3}$}
\email{lychkovskiy@gmail.com}

\affiliation{$^1$ Department of Mathematical Methods for Quantum Technologies, Steklov Mathematical Institute of Russian Academy of Sciences, 8 Gubkina St., Moscow 119991, Russia.}
\affiliation{$^2$ Russian Quantum Center, Skolkovo, Moscow 143025, Russia.}
\affiliation{$^3$ Skolkovo Institute of Science and Technology, Bolshoy Boulevard 30, bld. 1, Moscow 121205, Russia.}


\date{\today}

\begin{abstract}
We consider two one-dimensional quantum $XX$ magnets linked by a periodically driven quantum point contact (QPC). If magnets are initially polarized in opposite directions, one expects that a spin current through the QPC will establish.  It has been shown recently [Phys. Rev. B 103, L041405 (2021)] that, in fact, when the driving frequency exceeds a critical value, the current halts completely, the QPC being effectively insulating. Here we enquire how this picture is affected by quantum dephasing.  Our findings reveal that any non-zero dephasing restores the current.

\end{abstract}

\maketitle

\subsection*{Introduction}

The subject of transport through a driven quantum point contact (QPC) has traditionally attracted considerable attention. The prospect of controlling the macroscopic quantum state of the electron gas via an external time-dependent potential promises both practical applications \cite{beenakker2019deterministic,nayak2008non} and intriguing theoretical insights \cite{kouwenhoven1991quantized,nakajima2016topological,lohse2016thouless,levitov1996electron,ivanov1997coherent,keeling2006minimal,dubois2013minimal}. 

Previous research by one of the authors \cite{gamayun2021nonequilibrium} revealed a nonequilibrium phase transition \cite{marro2005nonequilibrium,prosen2011nonequilibrium} in a closed system consisting of two tight-binding free-fermionic chains separated by a periodically driven QPC. Specifically, it was found that when the driving frequency $\omega$ exceeds a critical value $\omega_c$ equal to the single-particle bandwidth of the chain, the interchain current drops to zero, i.e. the QPC becomes insulating.\footnote{In general, one expects that the quantum dynamics should be suppressed when the driving frequency exceeds the bandwidth. In fact,   one can prove that for locally interacting many-body systems this suppression is exponential in the frequency \cite{Abanin_2015}. The result of Ref. \cite{gamayun2021nonequilibrium} is, however, stronger: it asserts that, for certain (but not all) QPCs, the cycle-averaged current  is exactly zero (and not merely suppressed) in the nonequilibrium steady state for an arbitrary frequency above the critical one.} Conversely, when the driving frequency  is less than this critical value, $\omega<\omega_c$, the QPC becomes conducting and a non-zero current between the chains is established. 

In the present Letter, we examine how a weak interaction with the environment modifies this picture. Specifically, we consider the effect of markovian dephasing that can be treated by means of the Gorini-Kossakowski-Sudarshan-Lindblad (GKSL) equation. We find that any finite  dephasing suffice to make the QPC conducting even for $\omega>\omega_c$, thus eliminating the nonequilibrium phase transition.

It is known that a tight-binding free-fermionic chain can be mapped to the one-dimensional spin-$1/2$ $XX$ model by means of the Jordan-Wigner transformation~\cite{Lieb_Schultz_Mattis_1961}. We find it convenient here to work in the spin language. Instead of two tight-binding chains, we consider two $XX$ magnets. Initially the magnets are oppositely polarized. The particle current in the fermionic language is then substituted by the spin current that tends to level the polarisation bias. 

The spin (or qubit) language is  particularly convenient in the context of quantum simulation and computation. Recent advancements in noisy intermediate-scale quantum (NISQ) devices \cite{mi2022noise,zhu2022observation}, such as superconducting processors and cold atom arrays, already allow experimental studies of topics from quantum many-body physics. It would be interesting to implement the setup proposed in \cite{gamayun2021nonequilibrium} on one of the existing NISQ devices. The dynamics of the $XX$ model is known to be equivalent to the sequence of certain two-qubit quantum gates known as matchgates \cite{valiant2001quantum,terhal2002classical,jozsa2008matchgates}, further simplifying implementation within the framework of universal quantum computation. 

A typical NISQ device is subject to dephasing. Thus it is natural to enquire what effect the dephasing will have on the phenomenon found in \cite{gamayun2021nonequilibrium}. This consideration additionally justifies the subject of our study.

We tackle the problem by solving coupled GKSL equations in the Heisenberg representation.  In the case of the  $XX$ model with dephasing, the space of operators is known to be fragmented into dynamically decoupled subspaces of varying dimensionality \cite{znidaric2010exact,Shibata_2019,turkeshi2021diffusion,teretenkov2023exact}. This brings a huge simplification and allows us to  numerically treat relatively large systems and, thereby, to draw a reliable physical picture.

\subsection*{General setup}

A Markovian dissipative dynamics can be described by the GKSL equation in the Heisenberg representation~\cite{breuer2002theory},
\begin{align}
\label{GKSL}
\partial_t O_t=i[H,O_t]+\mathcal{D}^\dagger O_t,
\end{align}
with the initial conditions $O_{t=0}=O$. Here $O_t$ and $O$  are,  respectively, Heisenberg and Schr\"odinger representations of the observable $O$, $H$ is a Hamiltonian and $\mathcal{D}^\dagger$ is an ajoint dissipation superoperator that reads
\begin{align}
\label{D_def}
\mathcal{D}^\dagger O_t\equiv\gamma\sum\limits_j\left(l^\dagger_j O_tl_j-\frac{1}{2}\{l^\dagger_j l_j,O_t\}\right),
\end{align}
where $l_j$ are Linblad operators, $\gamma$ is a real positive constant and $\{\cdot,\cdot\}$ denotes an anticommutator. If the Heisenberg operator $O_t$ of an observable is known,  time evolution of its expectation value is given by $\langle O \rangle_t = {\rm tr}\, O_t \rho_0$, where $\rho_0$ is an initial state of the system.

The Hamiltonian of the system under consideration reads ({\it cf.} \cite{gamayun2021nonequilibrium}):

\begin{align}
\label{ham_tot}
H=H_\text{L}+H_\text{R}+V_t,
\end{align}
where $H_\text{L}$ and $H_\text{R}$ describe two XX magnets, and $V_t$ describes the driven QPC connecting these two magnets. Explicitly, 
\begin{align}
\label{ham_lr}
&H_\text{L}=\frac{1}{4}\sum\limits^{L-1}_{j=1}(\sigma^x_j\sigma^x_{j+1}+\sigma^y_j\sigma^y_{j+1}),\nonumber\\
&H_\text{R}=\frac{1}{4}\sum\limits^{2L-1}_{j=L}(\sigma^x_j\sigma^x_{j+1}+\sigma^y_j\sigma^y_{j+1}),
\end{align}
\begin{align}
\label{qpc_ham}
V_t=\frac{\sin(\omega t)}{4}(\sigma^x_L\sigma^x_{L+1}+\sigma^y_L\sigma^y_{L+1}),
\end{align}
where $\sigma^\alpha_j, \alpha=x,y,z$ are Pauli matrices at the $j$'th site, $L$ refers to the number of spins in each magnet, and $\omega$ is the driving frequency. Note that $V_t$ is the only term of the Hamiltonian that depends on time. $V_t$ vanishes in the limit of  $\omega=0$; in this limit the magnets become disconnected.

The Lindblad operators $l_j$ are given by
\begin{equation}\label{L}
l_j=\sigma^z_j,\quad j=1,2,\dots,2L.
\end{equation}
Such Lindblas opearators are known to cause dephasing, i.e. the  decay of off-diagonal elements of the density matrix in the $\sigma^z_j$ eigenbasis. 

Initially magnets are prepared in a pure state $\rho_0=|\Psi_0\rangle \langle\Psi_0|$, where 
\begin{align}
\label{psi_ini}
|\Psi_0\rangle=|0_1\dots 0_{L}\rangle\otimes|1_{L+1}\dots1_{2L}\rangle
\end{align}
and $|0_j\rangle$, $|1_j\rangle$ are eigenvectors of $\sigma^z_j$  such that $\sigma^z_j|0_j\rangle=-|0_j\rangle$,  $\sigma^z_j|1_j\rangle=|1_j\rangle$. 

The initial condition (\ref{psi_ini}) means that  left and right magnets are completely polarized in the opposite directions, see Fig. \ref{pic:magProfile}. Notably, in the limit of  $\omega=0$, i.e. when the the magnets are disconnected, this state is the eigenstate of the Hamiltonian (\ref{ham_tot}). Moreover, the corresponding density matrix $\rho_0$ is the steady state of the GKSL equation (\ref{GKSL}). Simply put, in the absence of QPC, the magnetization profile defined by (\ref{psi_ini}) remains unchanged over time, whether the dephasing is present or not.

\subsection*{Solving coupled GKSL equations}

Generally, the numerical solution of the GKSL equation (\ref{GKSL}) requires an exponential amount of resources. This is because the dimension of the space of operators for  $2L$ qubits grows as $4^{2L}$. However, for some dissipative systems the space of operators gets fragmented into dynamically disconnected sectors, with the dimension of some sectors being polynomial in the number of qubits \cite{prosen2008third,znidaric2010exact,eisler2011crossover,temme2012stochastic,vzunkovivc2014closed,Essler_2020,bakker2020lie,linowski2022dissipative}. The system under consideration is of this type \cite{znidaric2010exact,Shibata_2019,turkeshi2021diffusion,teretenkov2023exact}. Specifically, the subspace containing our observables of interest,  $z$ projections of spin polarizations,  $\sigma^z_j$, has the dimension that scales as $L^2$. Below we explicitly construct this subspace. 

First we consider the model without dissipation,  $\gamma=0$. In this case, the system is closed and equation~(\ref{GKSL}) is the Heisenberg equation. We introduce the following  operators known as Onsager strings ({\it cf.} \cite{jha1973xy,prosen1998new,lychkovskiy2021closed,teretenkov2023exact}):

\begin{align}
\label{a_and_b}
&A^0_j=-\sigma^z_j\nonumber\\
&A^{n}_j=\sigma^x_j\left(\prod\limits^{n-1}_{m=1}\sigma^z_{j+m}\right)\sigma^x_{j+n},\\
&A^{-n}_j=\sigma^y_j\left(\prod\limits^{n-1}_{m=1}\sigma^z_{j+m}\right)\sigma^y_{j+n},\nonumber\\
&B^{n}_j=\frac{i}{2}\sigma^x_j\left(\prod\limits^{n-1}_{m=1}\sigma^z_{j+m}\right)\sigma^y_{j+n},\nonumber\\
&B^{-n}_j=-\frac{i}{2}\sigma^y_j\left(\prod\limits^{n-1}_{m=1}\sigma^z_{j+m}\right)\sigma^x_{j+n},
\quad 1\leq n\leq 2L-1.\nonumber
\end{align}
Here $n+1$ is the ``size'' of an Onsager string, i.e. the number of Pauli matrices it contains. This size runs from one (for $A^0_j$) to $2L$ (for $A_1^{\pm(2L-1)}$, $B_1^{\pm(2L-1)}$). Note that index $j$ should be consistent with $n$: namely, $j=1,2,\dots,2L-n$ are allowed for a given $n$. This rule implies that there are $D=2L(4L-1)$ Onsager strings in total. 

It is easy to see that the operator subspace $\mathcal{P}$ spanned by these $D$ Onsager strings is closed with respect to commutation with the Hamiltonian~\eqref{ham_tot}~\cite{jha1973xy,prosen1998new,lychkovskiy2021closed}, as demonstrated explicitly in the Supplement \cite{supp}.  Thus this subspace is decoupled from the rest of the operator space under the evolution governed by the Heisenberg equation.

Let us now turn to the case with dissipation,  $\gamma>0$. It is easy to verify  that the subspace $\mathcal{P}$ is invariant under the dissipation superoperator  with Lindblad operators~\eqref{L} \cite{znidaric2010exact,Shibata_2019,turkeshi2021diffusion,teretenkov2023exact}. This follows from the equalities $\mathcal{D}\sigma^{x,y}_j=-2\sigma^{x,y}$ and $\mathcal{D}\sigma^{z}_j=0$ (see the Supplement \cite{supp} for more details).

As a consequence, a system of $D$ coupled  GKSL equations completely determines the dynamics within the subspace $\mathcal{P}$. Since $D$ is only quadratic in the system size, these equations can be  efficiently solved for relatively large system sizes $L$. This way we are able to numerically treat systems consisting of a few dozens of qubits on a laptop, obtaining the magnetization profile as a function of time. The results are presented in the next section.

Let us briefly outline  the fermionic picture of our setting. Under the Jordan-Wigner transformation~~\cite{Lieb_Schultz_Mattis_1961}, the Hamiltonian \eqref{ham_tot} describes two tight-binding noninteracting fermionic chains connected by a QPC with a periodically varying tunneling \cite{gamayun2021nonequilibrium}. A local spin operator $\sigma^z_j$ maps to  $2n_j-1$, where $n_j$ is the fermionic number operator on the $j$'th site, the conservation of the total $z$-magnetization corresponds to the particle number conservation, the spin current maps to the particle current and the initial state \eqref{psi_ini} corresponds to the left chain being empty and the right chain being completely filled by fermions. The Onsager strings~\eqref{a_and_b} are quadratic in fermionic creation and annihilation operators and span the subspace of all quadratic operators. 

The latter fact immediately explains the invariance of the space of Onsager strings under the purely coherent dynamics  generated by the  Hamiltonian \eqref{ham_tot} (which is also quadratic in the fermionic picture). 

The reason for the invariance in the presence of dissipation is more subtle. The dissipation superoperator with Lindblad operators \eqref{L} is not quadratic but fourth order \cite{znidaric2010exact,eisler2011crossover,temme2012stochastic}. One could argue, however, that these Lindblad operators are equivalent to the stochastic local magnetic fields (in the spin picture) or  chemical potentials (in the fermionic picture), see e.g. \cite{Kiely_2021}. This brings one back to a quadratic Hamiltonian, though with stochastic terms. This reasoning is, however, specific for particular Lindblad operators \eqref{L}. In fact, the aforementioned invariance emerges for broad classes of Lindblad operators that, in general, are not equivalent to quadratic stochastic Hamiltonians or  quadratic Lindbladians  \cite{teretenkov2023exact}. For example, this is the case for Lindblad operators $l_j=\sigma^z_j\sigma^z_{j+1},\quad j=1,\dots,2L-1$  that correspond  to fourth order terms in the corresponding stochastic Hamiltonian.  We have repeated our calculations for this set of Lindblad operators, see  the Supplement \cite{supp}. The results are qualitatively the same as for Lindblad operators \eqref{L}.

\subsection*{Results}

\begin{figure}
\includegraphics[width=\columnwidth]{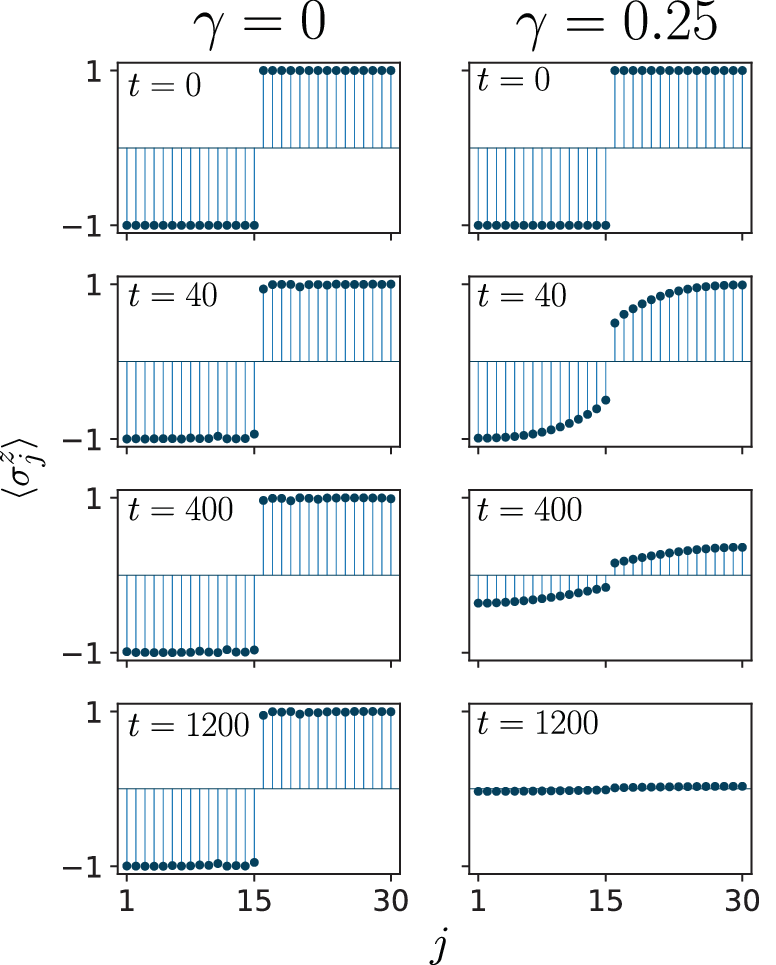}
\caption{
Fig. \ref{pic:magProfile} Snapshots of the magnetization profile of the two  $XX$ spin chains connected by the QPC and initialized in the state (\ref{psi_ini}), in the absence (left column) and presence (right column) of dephasing. The total number of spins is $2L=30$, the driving frequency is $\omega=2.5$. One can see that in the absence of dephasing the QPC is insulating, while in the presence of dephasing the QPC conducts the spin current.
\label{pic:magProfile}}
\end{figure}

It has been shown in  \cite{gamayun2021nonequilibrium} that, in the absence of dissipation, the QPC turns insulating for driving frequencies exceeding $\omega_c=2$. We start from verifying this fact using our approach. To this end we perform numerical simulations of the magnetization profile for $\gamma=0$ and $\omega=2.5$. The results are shown in the left panel of Fig.  \ref{pic:magProfile}. One can see that, apart from a small initial ``leak'' of magnetization occurring during the first few cycles (which is a transient effect also observed in \cite{gamayun2021nonequilibrium}), the QPC  indeed  preserves the initial magnetization imbalance. 

To confirm that the system has indeed essentially approached the  nonequilibrium steady state  within the studied timescale, we compute the following quantity:
\begin{align}
\label{del_r}
\Delta^R(t)=L-\sum\limits^{2L}_{j=L+1}\langle\sigma^z_j\rangle(t).
\end{align}
This quantity measures the deviation of the total magnetization of the right magnet from the initial magnetization. If the QPC conducts the spin current, then the magnetization (or, equivalently, polarization) vanishes  and $\Delta^R(t)\rightarrow L$ at $t\rightarrow \infty$. In contrast, if the QPC is insulating, $\Delta^R(t)$  should not grow with the system size. Instead, it swiftly approaches some  (typically, small) value that is finite in the limit of   $L\rightarrow\infty$. The latter behaviour  is a manifestation of the initial leak of magnetization.
\footnote{The nonzero value of this leak highlights the fact that the magnetization of either of the two magnets is {\it not} a conserved quantity (as it would be in the case of disconnected magnets), and the initial state \eqref{psi_ini} is {\it not} a steady state. Rather, the leak accompanies the relaxation of the initial state to the nonequilibrium steady state.}

\begin{figure}
\includegraphics[width=\columnwidth]{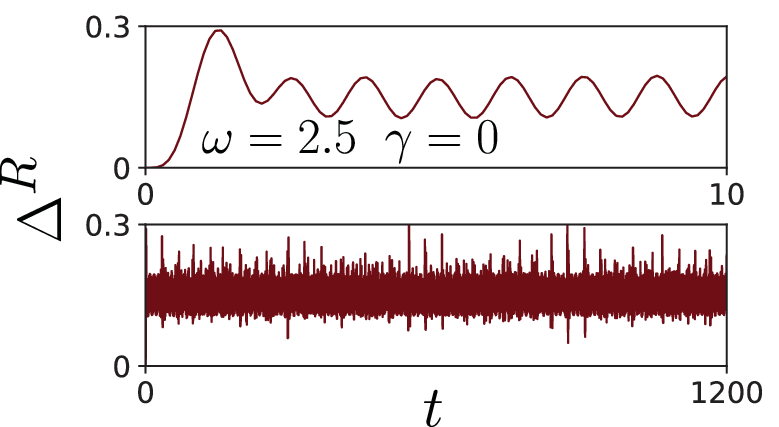}
\caption{
Fig. \ref{pic:gam0} Deviation $\Delta^R(t)$ of the total magnetization of the right chain from the initial magnetization, in the absence of dephasing. The top plot spans first few cycles of oscillations, the bottom one -- the whole timescale of Fig. \ref{pic:magProfile}. One can see that $\Delta^R(t)$ remains small  and does not show any tendency to approach $L=15$, which means that the QPC is insulating, consistent with the results of ref. \cite{gamayun2021nonequilibrium}.
\label{pic:gam0}}
\end{figure}

In Fig. \ref{pic:gam0} we demonstrate  that, in the case of no dephasing,  $\Delta^R(t)$ remains below 1 and does not show any tendency to approach $L$. We average  $\Delta^R(t)$ 
 over time to obtain $\langle\Delta^R\rangle^{2L=30}_\text{mean}=0.150$, with the root mean square value $\langle\Delta^R\rangle^{2L=30}_\text{rms}=0.155$. We also verify that this value does not grow with the system size, in particular, $\langle\Delta^R\rangle^{2L=20}_\text{mean}=0.151$ and $\langle\Delta^R\rangle^{2L=10}_\text{mean}=0.156$ (with  $\langle\Delta^R\rangle^{2L=20}_\text{rms}=0.150$, $\langle\Delta^R\rangle^{2L=10}_\text{rms}=0.160$). We therefore conclude that the QPC is indeed insulating. 

Then we perform calculations for non-zero dephasing~$\gamma$. We find that in this case  the QPC is always conductive, as illustrated in the right column of Fig. \ref{pic:magProfile}. In this case, the magnetization imbalance is levelled with time,   the left and right parts of the system eventually becoming completely depolarized.  

\begin{figure}
\includegraphics[width=\columnwidth]{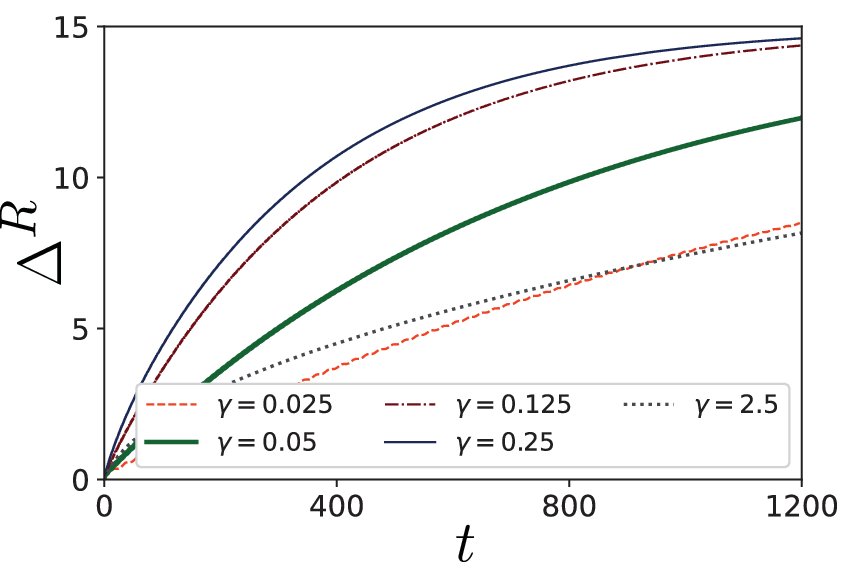}
\caption{
Fig. \ref{pic:deltaM} Deviation $\Delta^R(t)$ of the total magnetization of the right chain from the initial magnetization for various values of dephasing $\gamma$. Total number of spins is $2L=30$, the driving frequency is $\omega=2.5$.
\label{pic:deltaM}}
\end{figure}

Fig. \ref{pic:deltaM} displays how $\Delta^R$ increases and eventually saturates at the value $L$ in the case of non-zero dephasing. The QPC becomes conductive even when dephasing is relatively small.

Interestingly, the current as a function of the dephasing strength  is nonmonotonic. For example, the growth rate of $\Delta^R$ for  $\gamma=2.5$ is smaller than for  $\gamma=0.25$, as illustrated in Fig. \ref{pic:deltaM}. This behavior is a manifestation of the  dissipative quantum Zeno effect \cite{misra1977zeno,presilla1996measurement}, where high dephasing effectively freezes the dynamics of the non-equilibrium state. Thus, the initial state (\ref{psi_ini}) is stable in the opposite limits of $\gamma=0$ and $\gamma\rightarrow\infty$.

\subsection*{Summary}

We have investigated the out-of-equilibrium physics of a system of two dissipative $XX$ magnets connected by a periodically driven quantum point contact. In the absence of dissipation, the contact was known to be non-conductive for frequencies above the critical one \cite{gamayun2021nonequilibrium}.  We demonstrate that this effect does not tolerate  dephasing -- the contact invariably becomes conductive when the dephasing is introduced.

\subsection*{Acknowledgements}
This work was funded by Russian Federation represented by the Ministry of Science and Higher Education (grant number 075-15-2020-788).

\bibliography{ref}

\end{document}